\documentclass{emulateapj}
\usepackage{mathptmx}
\usepackage[latin9]{inputenc}
\usepackage{color}
\usepackage{booktabs}
\usepackage{textcomp}
\usepackage{url}
\usepackage{amstext}
\usepackage{amssymb}
\usepackage{graphicx}
\usepackage[unicode=true,
 bookmarks=false,
 breaklinks=false,pdfborder={0 0 0},backref=false,colorlinks=true]
 {hyperref}
\hypersetup{pdftitle={Star Formation in High-Redshift Cluster Ellipticals},
 pdfauthor={Wagner et al.},
 citecolor=blue, urlcolor=blue}
\usepackage{breakurl}

\makeatletter

\providecommand{\tabularnewline}{\\}

\newcommand{\hst}{{\it HST}}
\newcommand{\spitzer}{{\it Spitzer}}
\newcommand{\chandra}{{\it Chandra}}
\newcommand{\herschel}{{\it Herschel}}
\newcommand{\msun}{$M_\odot$} 
\newcommand{\lsun}{$L_\odot$} 
\newcommand{\UMKC}{1}
\newcommand{\queens}{2}
\newcommand{\STSCI}{3}
\newcommand{\florida}{4}
\newcommand{\ucdavis}{5}
\newcommand{\umass}{6}
\newcommand{\JPL}{7}
\newcommand{\penn}{8}
\newcommand{\SpitzerCenter}{9}
\newcommand{\NOAO}{10}
\newcommand{\siena}{11}

\altaffiltext{\UMKC}{Department of Physics and Astronomy, University of Missouri, 5110 Rockhill Road, Kansas City, MO 64110, USA}
\altaffiltext{\queens}{Current Affiliation: Department of Physics, Engineering Physics \& Astronomy, Queen's University, Stirling Hall, Kingston, ON K7L 3N6, Canada; cwagner@astro.queensu.ca}
\altaffiltext{\STSCI}{Space Telescope Science Institute, 3700 San Martin Drive, Baltimore, MD 21218, USA}
\altaffiltext{\florida}{Department of Astronomy, University of Florida, Gainesville, FL 32611, USA}
\altaffiltext{\ucdavis}{University of California, Davis, CA 95616, USA}
\altaffiltext{\umass}{Department of Astronomy, University of Massachusetts, Amherst, MA 01003, USA}
\altaffiltext{\JPL}{Jet Propulsion Laboratory, California Institute of Technology, Pasadena, CA 91109, USA}
\altaffiltext{\penn}{Department of Astronomy and Astrophysics, Pennsylvania State University, 525 Davey Laboratory, University Park, Pennsylvania 16802, USA}
\altaffiltext{\SpitzerCenter}{Spitzer Science Center, MC 220-6, California Institute of Technology, 1200 East California Boulevard, Pasadena, CA 91109, USA}
\altaffiltext{\NOAO}{NOAO, 950 North Cherry Avenue, Tucson, AZ 85719, USA}
\altaffiltext{\siena}{Department of Physics and Astronomy, Siena College, 515 Loudon Road, Loudonville, NY 12211, USA}

\shorttitle{Star Formation in High-Redshift Cluster ETGs}
\shortauthors{Wagner et al.}

\makeatother

\begin{document}
\begin{abstract}
We measure the star formation rates (SFRs) of massive ($M_{\star}>10^{10.1}$
\msun) early-type galaxies (ETGs) in a sample of 11 high-redshift
($1.0<z<1.5$) galaxy clusters drawn from the IRAC Shallow Cluster
Survey (ISCS). We identify ETGs visually from \textit{Hubble Space
Telescope} imaging and select likely cluster members as having either
an appropriate spectroscopic redshift or red sequence color. Mid-infrared
SFRs are measured using \spitzer\ 24 $\mu$m data for isolated cluster
galaxies for which contamination by neighbors, and active galactic
nuclei, can be ruled out. Cluster ETGs show enhanced specific star
formation rates (sSFRs) compared to cluster galaxies in the local
Universe, but have sSFRs more than four times lower than that of field
ETGs at $1<z<1.5$. Relative to the late-type cluster population,
isolated ETGs show substantially quenched mean SFRs, yet still contribute
12\% of the overall star formation activity measured in $1<z<1.5$
clusters. We find that new ETGs are likely being formed in ISCS clusters;
the fraction of cluster galaxies identified as ETGs increases from
34\% to 56\% from $z\sim1.5\rightarrow1.25$. While the fraction of
cluster ETGs that are highly star-forming ($\textrm{SFR}\geq26$ \msun\
yr$^{-1}$) drops from 27\% to 10\% over the same period, their sSFRs
are roughly constant. All these factors taken together suggest that,
particularly at $z\gtrsim1.25$, the events that created these distant
cluster ETGs---likely mergers, at least among the most massive---were
both recent and gas-rich.

\end{abstract}

\title{Star Formation in High-Redshift Cluster Ellipticals}

\author{Cory R. Wagner\altaffilmark{\UMKC}$^{,}$\altaffilmark{\queens},
Mark Brodwin\altaffilmark{\UMKC}, Gregory F. Snyder\altaffilmark{\STSCI},
Anthony H. Gonzalez\altaffilmark{\florida}, S. A. Stanford\altaffilmark{\ucdavis},
Stacey Alberts\altaffilmark{\umass}, Alexandra Pope\altaffilmark{\umass},
Daniel Stern\altaffilmark{\JPL}, Gregory R. Zeimann\altaffilmark{\penn},
Ranga-Ram Chary\altaffilmark{\SpitzerCenter}, Arjun Dey\altaffilmark{\NOAO},
Peter R. M. Eisenhardt\altaffilmark{\JPL}, Conor L. Mancone\altaffilmark{\florida},
and John Moustakas\altaffilmark{\siena}}

\maketitle

\section{Introduction}

\label{Sec: Introduction}

In the local Universe galaxy clusters are primarily populated by quiescent,
early-type galaxies (ETGs) with little ongoing star formation and
evolved stellar populations \citep{oemler1974,dressler1980,caldwell1993,gomez2003,bressan2006,clemens2009,edwards2011}.
Studies of their galaxy populations to $z\lesssim1$ find that the
evolution in the color and scatter of their red-sequences are consistent
with simple passive evolution models in which the bulk of their stars
formed in a short, high-redshift starburst \citep{bower1992,aragon-salamanca1993,stanford1998,kodama1999,blakeslee2006,mei2006,mei2009,eisenhardt2008,muzzin2008}.
However, $\Lambda$CDM predicts a more extended, hierarchical formation
history. For instance, simulations by \citet{delucia2006} find that
only $\sim$50\% of massive elliptical galaxies would have formed
the bulk (80\%) of their stellar mass by $z\sim1.5$. A number of
studies in the $z\lesssim1$ regime have found that the star formation
of cluster galaxies increases with redshift \citep{couch1987,saintonge2008,finn2008,webb2013}.
Recently, infrared (IR) measurements of the $z>1$ cluster population
have revealed substantial dust-obscured star formation activity \citep[hereafter B13]{hilton2010,tran2010,santos2013,brodwin2013}.
\citet[hereafter A14]{alberts2014} found that dust-obscured star
formation in cluster galaxies increases with lookback time from $z=0.3\rightarrow1.5$.

Several recent studies of the high-redshift ($z>1$) cluster population
have been conducted using the IRAC Shallow Cluster Survey \citep[ISCS;][]{eisenhardt2008}.
\citet{mancone2010,mancone2012} measured the rest-frame near-infrared
(NIR) luminosity function evolution, and found that while it matched
what would be expected from passive evolution up to $z\sim1.3$, it
disagreed with such a model at $z\gtrsim1.3$, which they suggested
as evidence for a significant epoch of galaxy assembly via merging
\citep[see, however,][]{andreon2013,wylezalek2014}. In the hierarchical
evolution framework, massive ETGs are formed as the result of major
mergers \citep{negroponte1983,barnes1988,naab2003,cox2006}, and these
mergers can cause bursts of star formation \citep{sanders1988,barnes1996,hopkins2008}
and fuel an active galactic nucleus \citep[AGN;][]{springel2005}
that quenches star formation on the order of a few 100 Myr \citep{dimatteo2005,hopkins2006}.
\citet[hereafter S12]{snyder2012} found that red-sequence members
have roughly constant stellar ages across $1.0<z<1.5$---indicating
that star formation must be ongoing---and bluer and more stochastic
colors at $1.0<z<1.3$ than would be expected of a passive population.
\citet{zeimann2013}, B13, and A14 measured high, and consistent,
H$\alpha$, 24 $\mu$m, and 250 $\mu$m star formation rates (SFRs),
respectively, in these clusters.

With these high SFRs, it is clear that members in $1<z<1.5$ clusters
have not exhausted their supplies of cold gas, which, in combination
with the evidence of significant ongoing merger activity, suggests
that a substantial number of these mergers are gas-rich. If gas-rich
major mergers are common in clusters then the ETGs formed in these
mergers would be expected to have high SFRs, at least for a short
time after their formation. Conversely, ETGs observed several hundred
Myr post-merger would likely appear to be recently quenched.

In this paper we use high-resolution \textit{Hubble Space Telescope}
(\hst) images of ISCS galaxy clusters at $1<z<1.5$ to identify isolated,
early-type members, and then measure their dust-inferred SFRs using
\spitzer\ 24 $\mu$m data. The goal of this work is to test whether
the high SFRs seen in high-redshift clusters are merely due to the
morphological mix---a result of the elevated late-type galaxy (LTG)
fraction relative to local clusters---or whether significant star
formation is present in the early-types as well.

In §\ref{Sec:Data} we summarize the ISCS cluster sample, multi-wavelength
data sets, and red-sequence catalogs of S12 which form the basis of
this work. In §\ref{Sec:Method} we describe our criteria for selecting
isolated, early-type cluster members, and present our measurements
of their star formation activity in §\ref{sec:Analysis}. We discuss
our results in §\ref{Sec:Discussion} and present our conclusions
in §\ref{Sec:Conclusions}. Throughout this work, we use AB magnitudes
unless otherwise indicated. Stellar masses are estimated using a \citet{chabrier2003}
initial mass function (IMF). We adopt a WMAP7 cosmology \citep{komatsu2011},
with ($\Omega_{\Lambda}$, $\Omega_{M}$, $h$) = (0.728, 0.272, 0.704).

\section{Data}

\label{Sec:Data}

\subsection{IRAC Shallow Cluster Survey}

The ISCS \citep{eisenhardt2008} identified candidate galaxy clusters
in a 7.25 deg$^{2}$ area of the Bo\"otes field of the NOAO Deep
Wide-Field Survey \citep[NDWFS;][]{jannuzi1999}, over a redshift
range of $0.1<z<2$, using imaging from the IRAC Shallow Survey \citep[ISS;][]{eisenhardt2004}.
Using accurate photometric redshifts from \citet{brodwin2006}, a
wavelet algorithm was used to identify clusters from the $4.5\,\mu$m-selected
galaxies as three-dimensional overdensities. The cluster centers were
taken to be the peaks in the wavelet detection maps. Three more epochs
were subsequently obtained in all IRAC bands as part of the \spitzer\
Deep, Wide-Field Survey \citep[SDWFS;][]{ashby2009}, which increased
the photometric depth of the IRAC images by a factor of two. The deeper
SDWFS data were used to improve the photometric redshift accuracy
for all galaxies, as well as to extend the catalog to lower flux limits.
The SDWFS catalog is 80\% complete at 18.1 mag in the 4.5$\mu$m band
\citep{ashby2009}.

In this work we focus on 11 spectroscopically confirmed $1<z<1.5$
ISCS clusters selected by S12 for follow-up \hst\ observations, as
described below. We list these clusters, along with their positions
and spectroscopic redshifts, in Table\ \ref{tab:clusterList}.

\begin{table}[!t]
\protect\caption{ISCS Clusters \label{tab:clusterList}}

\begin{centering}
\begin{tabular}{cccccc}
\addlinespace[-0.02\textheight]
 &  &  &  &  & \tabularnewline
\midrule
\midrule 
ISCS Cluster & R.A. & Dec. & $z_{\textrm{spec}}$ & $N_{\textrm{ETG}}$%
\footnote{Number of visually selected ETGs%
} & $N_{\textrm{LTG}}$%
\footnote{Number of visually selected LTGs%
}\tabularnewline
Name & (J2000) & (J2000) &  &  & \tabularnewline
\midrule 
J1429.2+3357 & 14:29:15.16 & 33:57:08.5 & 1.059 & 8 & 5\tabularnewline
J1432.4+3332 & 14:32:29.18 & 33:32:36.0 & 1.112 & 3 & 4\tabularnewline
J1426.1+3403 & 14:26:09.51 & 34:03:41.1 & 1.136 & 5 & 7\tabularnewline
J1426.5+3339 & 14:26:30.42 & 33:39:33.2 & 1.163 & 8 & 5\tabularnewline
J1434.5+3427 & 14:34:30.44 & 34:27:12.3 & 1.238 & 3 & 4\tabularnewline
J1429.3+3437 & 14:29:18.51 & 34:37:25.8 & 1.262 & 6 & 3\tabularnewline
J1432.6+3436 & 14:32:38.38 & 34:36:49.0 & 1.349 & 3 & 4\tabularnewline
J1433.8+3325 & 14:33:51.13 & 33:25:51.1 & 1.369 & 5 & 7\tabularnewline
J1434.7+3519 & 14:34:46.33 & 35:19:33.5 & 1.372 & 2 & 7\tabularnewline
J1438.1+3414 & 14:38:08.71 & 34:14:19.2 & 1.414 & 4 & 9\tabularnewline
J1432.4+3250 & 14:32:24.16 & 32:50:03.7 & 1.487 & 4 & 6\tabularnewline
\midrule
\multicolumn{4}{r}{Total Number} & 51 & 61\tabularnewline
\bottomrule
\end{tabular}
\par\end{centering}

\textbf{Notes.}
\end{table}

\subsection{\hst\ Data}

\label{sub:HST_data}

A subset of ISCS clusters spanning $1<z<1.5$ were imaged with \hst\
in the NIR and optical with instrument and filter combinations chosen
to bracket the 4000 \AA \ break. NIR data were acquired with the
Wide Field Camera 3 \citep[WFC3;][]{kimble2008} at F160W. In the
optical, observations were either taken with the Advanced Camera for
Surveys \citep[ACS;][]{ford1998} in filters F775W, F814W or F850LP,
or with the Wide Field Planetary Camera 2 \citep[WFPC2;][]{holtzman1995}
at F814W. The reader is referred to S12 for a more detailed description
of these data.

We use the all-\hst\ color-magnitude relations (CMRs) of 11 distant,
spectroscopically confirmed ISCS clusters presented in S12. That work
isolated the CMRs by subtracting a passively evolving Coma CMR model
with a fixed rest-frame slope. Galaxies brighter than an evolving
magnitude limit of $H^{*}(z)+1.5$ with color offsets, $\Delta$,
within $-0.25<\Delta<0.75$ of the model CMR were identified as red-sequence
galaxies. To reduce the effect of outliers, they removed galaxies
that were more than two median absolute deviations in color from their
measured $\Delta$ zero point.

We independently calculate the F160W completeness via Monte Carlo
methods. We insert 100 artificial stars into each WFC3 cluster image
in steps of 0.05 magnitudes from 23 to 25.5, ensuring a minimum separation
of 5 pixels between input stars. We repeat this procedure 10 times
providing 1000 input sources per magnitude bin, per cluster. We run
\textsc{SExtractor }\citep[ver. 2.8.6,][]{bertin1996} on each image
to generate source positions and \textsc{mag auto} magnitudes. We
match the \textsc{SExtractor} positions with our input positions in
each magnitude bin, counting the source as matched if it lies within
2 pixels of the input position of the artificial star. For each matched
source, we calculate the magnitude difference between the input source
and \textsc{mag auto} output, and find that the absolute mean $\Delta$mag
is $\lesssim0.2$ up to a magnitude of 24.9. We perform a least squares
fit to the fraction of matched sources and find a 90\% completeness
limit of 24.1 mag, consistent with that found in S12. We confirm that
all galaxies in our final cluster sample (§\ref{Sec:Method}) have
an F160W magnitude brighter than our 90\% completeness.

\subsection{Mid-Infrared Data}

\label{sub:midInfrared_data}

The high-redshift clusters studied in this work were also imaged at
$24$ $\mu$m with the Multiband Imaging Photometer for \spitzer\
\citep[MIPS;][]{rieke2004}. The exposure times, which increased with
redshift from 12 to 48 min, were designed to produce similar sensitivities
in IR luminosity for all clusters. Following the method of \citet{magnelli2009},
MIPS source catalogs were generated by using the positions of objects
in the higher-resolution IRAC images as priors. This method produces
24 $\mu$m flux measurements (or limits) for all IRAC galaxies. For
consistency with B13, we infer total IR luminosities for these sources
using templates from \citet{chary2001}, and convert these to SFRs
using the \citet{murphy2011} relation. The 1 $\sigma$ depth of our
SFRs is $\sim$ 13 \msun\ yr$^{-1}$.

While the \citet{chary2001} templates typically overestimate $L_{\textrm{IR}}$
at $z>1.5$ by a factor of $\sim$2--8 \citep{murphy2009,nordon2010,rodighiero2010},
it provides an estimate that is accurate to 40\% up to $z\sim1.5$
\citep{marcillac2006,murphy2009,elbaz2010}. 

In determining their total $L_{\textrm{IR}}$ to SFR calibration,
\citet{murphy2011} assumed that the entire Balmer continuum is absorbed
and reradiated by optically thin dust. They also assumed a solar metallicity,
and continuous star formation over a timescale of $\sim$100 Myr.
The relation was defined using a \citet{kroupa2001} IMF, which has
a similar normalization to the \citet{chabrier2003} IMF we use to
calculate our stellar masses.

\subsection{Chandra X-ray Data}

\citet{murray2005} obtained X-ray imaging of the NDWFS field with
\chandra\ to depths of 5--15 ks. Follow-up Cycle 10 \chandra\ observations
brought the exposure time to a uniform depth of 40 ks for the clusters
in the present sample.

\subsection{Comparison Field Sample Data}

We select our comparison sample of field galaxies from UltraVISTA
\citep{mccracken2012,muzzin2013}, a deep $K_{s}$-selected survey
covering 1.62 deg$^{2}$ of the COSMOS field \citep{scoville2007}.
The publicly available UltraVISTA survey%
\footnote{\url{http://www.strw.leidenuniv.nl/galaxyevolution/ULTRAVISTA/Ultravista/Data_Products_Download.html}%
} has photometry for $\sim$260,000 sources in 30 bands, including
\spitzer\ photometry in the 24\ $\mu$m band of MIPS and all four
channels of IRAC \citep{muzzin2013}.

We prune the catalog of stellar objects, contamination from bright
stars, and contamination from nearby saturated objects, and select
only galaxies brighter than the UltraVISTA 90\% completeness limit
of $K_{s,\textrm{tot}}=23.4$.

\citet{muzzin2013} estimated stellar masses by using \textsc{FAST}
\citep{kriek2009} to fit galaxy spectral energy distributions (SEDs)
to template SEDs. The templates were generated with \citet{bruzual2003}
models, using a \citet{chabrier2003} IMF. Photometric redshifts in
UltraVISTA were derived by \citet{muzzin2013} using \textsc{EAZY}
\citep{brammer2008}. A small subset of our final sample ($N=6$)
has spectroscopic redshifts from the zCOSMOS surveys \citep{lilly2007,lilly2009}.

Following the method in §\ref{sub:midInfrared_data}, we infer total
IR luminosities for UltraVISTA galaxies using the \citet{chary2001}
templates, and calculate SFRs with the \citet{murphy2011} relation.

\section{Galaxy Selection Method}

\label{Sec:Method}

We first build our cluster member sample by selecting galaxies that
are robust spectroscopic or red-sequence members (§\ref{sub:cluster_members}).
We remove galaxies that likely harbor AGNs (§\ref{sub:AGN_rejection}),
and galaxies that fall below our uniform stellar mass cut (§\ref{sub:stellar_mass}).
We then describe our selection of a comparison sample of high-redshift
field galaxies in §\ref{sub:FieldComp}. In order to robustly measure
SFRs for cluster members, we select galaxies that are free from potential
24\ $\mu$m contamination due to nearby neighbors, based on visual
inspection of optical and IR images (§\ref{sub:Isolation}). We then
visually classify our isolated cluster members, separating them into
ETGs and LTGs (§\ref{sub:morph}).

\subsection{Identification of Cluster Members}

\label{sub:cluster_members}

We select the 996 galaxies from the 11 high-redshift ISCS clusters
studied in S12 for which we have optical/NIR (\hst) and 24 $\mu$m
(\spitzer) images. We match these galaxies to the 8683 sources in
the IRAC 4.5 $\mu$m-selected SDWFS catalog, choosing galaxies where
the separation between the \spitzer\ and \hst\ positions is $\leq$
2\arcsec. This leaves 505 position matched galaxies with \hst\ imaging,
and SFR and stellar mass estimates.

SDWFS is effectively a mass-selected catalog, and a portion of the
491 \hst-detected galaxies that are not matched in SDWFS fall below
the uniform IRAC-based stellar mass cut we impose below. The remainder
of the unmatched S12 catalog members are likely in close proximity---at
least in projection---to some of the matched galaxies. However, in
§\ref{sub:Isolation}, we remove such non-isolated galaxies. Our F160W
images cover roughly the central 2\arcmin $\times$ 2\arcmin\ of
the $\sim$10\arcmin $\times$ 10\arcmin\ IRAC 4.5 $\mu$m SDWFS
images. As such, the vast majority of the 8178 unmatched SDWFS sources
lie outside of the HST images. However, we estimate that a few hundred
of these unmatched galaxies are undetected in F160W.

To identify the subset of galaxies that are robust cluster members,
we only retain for the final catalog objects with either high-quality
spectroscopic redshifts consistent with membership (\citealt{eisenhardt2008};
B13; \citealt{zeimann2013}), or those that are red-sequence members
based on \hst\ photometry, as described in §\ref{sub:HST_data}.
We cut 231 non-members, reducing our sample to 274 galaxies.

Galaxies that are heavily star-forming might exhibit colors outside
the range expected of a galaxy on the red sequence. If such galaxies
are indeed members, but also lack spectroscopic redshifts, then they
will be excluded from our sample. As such, our analysis yields a lower
bound on the total SFR. To assess the impact of this effect, we consider
our spectroscopic completeness.

We find that of the 505 position matched galaxies identified above,
252 are classified as non-members through S12's red-sequence analysis.
Of these, 43 (17\%) have high-quality spectra. However, this does
not include 71 galaxies that also have high-quality spectroscopic
redshifts, yet were confirmed as non-members, and not included in
S12's analysis. Hence, we do not have color information for them,
nor are they part of the position matched subset. They do, however,
still provide an upper limit of 35\% (114/323) to the spectroscopic
completeness of galaxies that would be excluded based on their colors.

We further consider morphology%
\footnote{We note that our use of morphology in this section is based on the
classifications we discuss in §\ref{sub:morph}.%
}, and find that both ETG and LTG red-sequence non-members have spectroscopic
completeness of 17\% when excluding the spectroscopically confirmed
non-members. If we include these galaxies, both morphologies have
an upper limit of 30\% completeness. The range of possible completeness
is fairly low, yet the same for both morphologies. We may indeed be
excluding a substantial number of galaxies, and while LTGs are the
likeliest candidates for the extreme SFRs mentioned above, they are
not likely preferentially excluded.

\subsection{Rejection of AGNs}

\label{sub:AGN_rejection}

The presence of an AGN can affect the MIR flux, potentially leading
to an incorrect estimate of the SFR. While only 1\% of local cluster
galaxies show AGN signatures \citep{dressler1985}, the surface density
of AGNs increases with redshift \citep{galametz2009,martini2013},
making the effect more prominent in our redshift range of interest.

Following B13, we remove AGNs identified via either X-ray or mid-IR
techniques. Galaxies whose counterparts in our 40 ks \chandra\ images
were point sources with hard X-ray luminosities brighter than $L_{X,H}>10^{43}$
erg s$^{-1}$ were removed as likely AGN. Similarly, objects with
signal-to-noise (S/N) $\geq5$ in all IRAC bands that fall in the
AGN wedge from \citet{stern2005}, which are reddened in the mid-IR
due to heating of their dust by AGN, were also removed. These cuts
removed a total 12 objects ($\sim$4\% of cluster members), bringing
our sample to 262 position-matched cluster members apparently free
of significant AGN contamination. The resulting SFRs will necessarily
be lower limits due to these exclusions.

\subsection{Stellar Masses and Mass Limit}

\label{sub:stellar_mass}

We measure stellar masses for our galaxies using the Bayesian SED
fitting code, \textsc{iSEDfit} \citep{moustakas2013}, which infers
galaxies' physical properties by fitting population synthesis models
to their broadband SEDs. In this work, we use population synthesis
models from \citet{bruzual2003}, which are based on the Padova 1994
stellar evolutionary tracks \citep{girardi1996}, the \textsc{stelib}
empirical stellar library \citep{leborgne2003}, and the \citet{chabrier2003}
IMF.

While all of the cluster members, and indeed all the galaxies from
our initial sample, are brighter than our 90\% F160W completeness
limit of 24.1 mag, we impose the $80$\% IRAC-based completeness limit
from B13. We remove the four galaxies that fall below this limit,
resulting in 258 cluster members with log($M_{\star}$$/$\msun)
$>10.1$.

\subsection{Comparison Field Sample Selection}

\label{sub:FieldComp}

To match the redshift range of our cluster sample, we select field
sources between $1<z<1.5$, based on the best redshift available.
While \citet{muzzin2013} reduced the effects of blending in their
24 $\mu$m photometry, to be consistent with our isolated criterion
for cluster galaxies, we select only isolated field objects by removing
photometric catalog members with a neighbor within 6\arcsec. We also
exclude likely AGNs using IRAC photometry to identify sources that
fall into the \citet{stern2005} wedge. We impose on our field sample
the same $80$\% IRAC-based completeness limit of log($M_{\star}$$/$\msun)
$>10.1$ that we use on our cluster galaxies. With these cuts, our
final field sample consists of 1127 isolated galaxies.

We use the updated, higher-redshift version of the morphological catalog
of \citet{cassata2007}%
\footnote{Publicly available COSMOS data sets, including the morphological catalog
used in this work, are located at \url{http://irsa.ipac.caltech.edu/data/COSMOS/datasets.html}%
} to separate ETGs and LTGs. They used a non-parametric automatic technique
to classify galaxies, based on the method of \citet{cassata2005},
who used concentration, asymmetry, and clumpiness parameters \citep{conselice2003,abraham2003,lotz2004}.
\citet{cassata2007} then extended this to include Gini, and M$_{20}$,
two concentration parameters.

While \citet{cassata2007} used an automatic technique to separate
galaxies, their classifications separate early- and late-type galaxies,
matching that of our classifications, which we will discuss below
in §\ref{sub:morph}. With the \citet{cassata2007} classifications,
our final sample contains 72 ETGs and 1055 LTGs.

\subsection{Visual Inspection}

\label{sub:visual_inspection}

\subsubsection{Isolation}

\label{sub:Isolation}

Because of its broad point spread function \citep[$\sim$6$\arcsec$,][]{rieke2004},
a single source in the $24$ $\mu$m band of MIPS can be comprised
of multiple distinct physical sources. Due to the difficulty in deconvolving
a multiple-object SFR into its constituent SFRs, we choose to limit
our work to isolated objects. To that end, we visually inspect $24$
$\mu$m \spitzer\ images and the available optical \hst\ images
of the 258 cluster members, removing from our final sample those for
which we are not able to rule out significant contributions to the
MIPS flux from other nearby objects. Through these inspections, we
further reduce our sample by 146 galaxies, resulting in 112 isolated
cluster members.

This cut on isolation removes interacting and merging members, thus
likely lowering our total measured SFRs. It was necessary, however,
to obtain robust SFR measurements for morphological early-types, a
key goal of this work. The significant star formation activity seen
amongst the isolated red-sequence galaxies (§\ref{sec:Analysis})
is therefore a lower limit. We do not attempt to correct the SFR to
the total value, but simply note that the sense of the correction---to
higher cluster SFRs for both early- and late-type members---serves
to strengthen our conclusions.

\subsubsection{Morphology}

\label{sub:morph}

\begin{figure}[!t]
\begin{centering}
\includegraphics{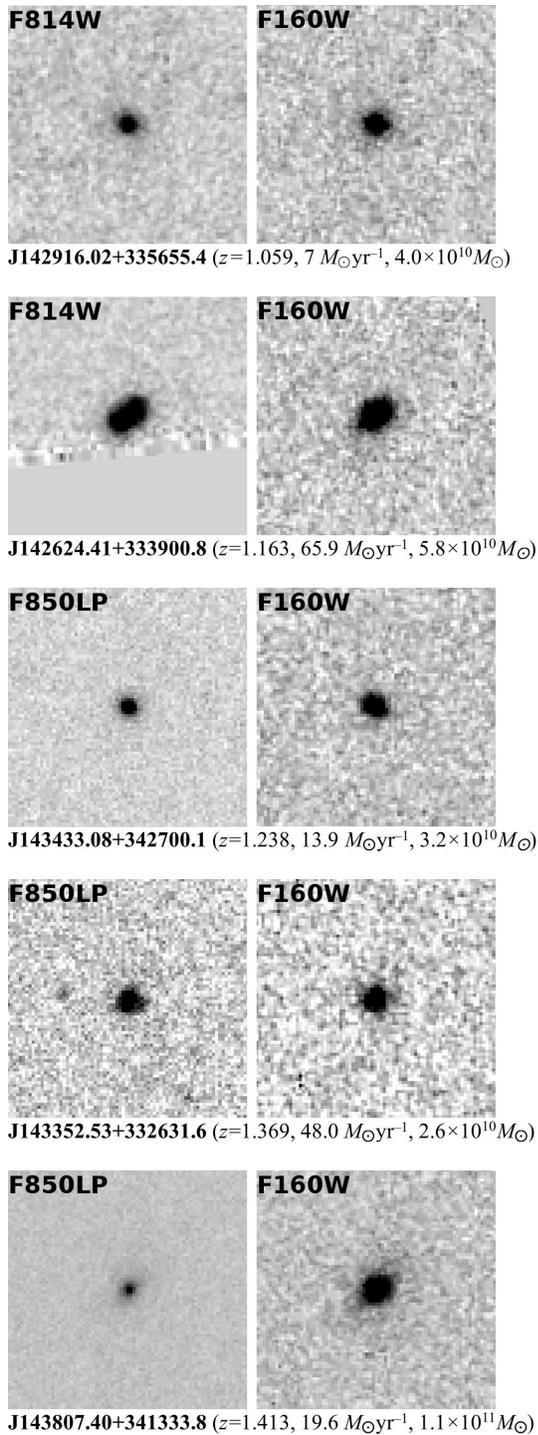}
\par\end{centering}

\protect\caption{5\arcsec$\times$5\arcsec\ cutouts of five isolated cluster ETGs,
with \hst\ filter listed on each image. Listed below each pair of
images is the galaxy name, spectroscopic cluster redshift, SFR, and
stellar mass. \label{fig:Cutouts}}
\end{figure}

We visually inspect the optical and NIR \hst\ images of the 112 isolated
galaxies in our sample, classifying those consistent with smooth elliptical
and S0 shapes as ETGs. Galaxies which exhibit either late-type signatures,
or disturbed or irregular morphologies, are collectively classified
as LTGs. By requiring the ETGs to have smooth early-type profiles
with no signs of interaction, we are removing from this sample galaxies
with merger signatures, which again biases our sample against ETGs
with potentially higher SFRs. In Figure\ \ref{fig:Cutouts}, we show
5\arcsec$\times$5\arcsec\ optical (left) and NIR (right) cutouts
of isolated cluster galaxies we classify as ETGs of varying SFRs and
redshifts. Below each galaxy, we list its name, spectroscopic cluster
redshift, SFR, and stellar mass. At these redshifts, 5\arcsec corresponds
to $\sim$41--43 kpc.

S12 performed visual inspections of their sample using F160W images,
assigning morphologies and recording the local environments. Additionally,
Sérsic indices ($n_{s}$) were measured in the F160W filter for all
S12 galaxies using \textsc{Galfit} \citep{peng2010} and G\textsc{alapagos}
\citep{haussler2011}, with ETGs defined as having $n_{s}>2.5$. These
Sérsic index measurements will be described in more detail by C. Mancone
et al. (2014, in preparation).

As a test of the robustness of our visual morphological classifications,
we compare with the independent S12 classifications. The primary difference
between these two morphological catalogs is that our classification
takes advantage of the higher resolution ACS and WFPC2 \hst\ images,
in addition to using the F160W images. Where the visual morphologies
differ, the cause tends to be late-type features (typically disks
with spiral structure) that are clearly visible in the high resolution
optical images but not apparent at F160W. Overall, we find agreement
for 91\% of the sample. We also test the Sérsic indices measured from
the F160W images, finding that this quantitative measure disagrees
with 31\% of our visual classifications. This high discrepancy level
is in line with the 30-40\% sample contamination reported by \citet{mei2012}
for selecting morphology using Sérsic indices.

We also test the consistency of using different classification methods
for galaxies in clusters and the field by visually classifying a subset
of 333 randomly selected UltraVISTA galaxies. We find that our classifications,
using COSMOS F814W images, agree for 294 galaxies (88\%).

\begin{figure}[!t]
\begin{centering}
\includegraphics{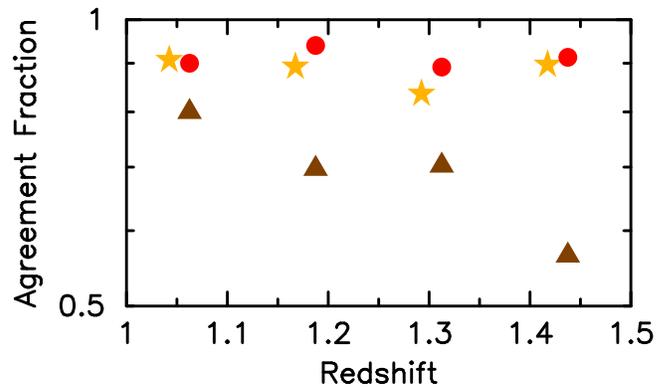}
\par\end{centering}

\protect\caption{Classification agreement for the isolated ISCS cluster sample, between
our visual classifications and S12's F160W-based visual (Sérsic) classifications,
shown by the red circles (brown triangles). The agreement between
our visual classifications and the \citet{cassata2007} quantitative
classifications for a test set of isolated UltraVISTA field galaxies
over $1<z<1.5$ is shown by the gold stars. Both the visual/visual
cluster agreement, and the visual/quantitative field agreement are
high, and show little-to-no redshift dependence. The visual/Sérsic
agreement, however, is uniformly lower, and negatively correlated
with redshift. \label{fig:morphAgreement}}
\end{figure}

Separating our galaxies into four evenly-sized bins, we plot in Figure\ \ref{fig:morphAgreement}
the visual/visual and visual/Sérsic agreement for cluster galaxies
as a function of redshift, as the red circles and brown triangles,
respectively. The agreement for the UltraVISTA test set is plotted
with the gold stars. The visual/visual agreement for cluster galaxies
is high, with no strong redshift dependence. The visual/quantitative
agreement for the field sample shows similar results. The cluster
visual/Sérsic agreement, however, is uniformly lower, and shows a
strong negative trend with increasing redshift.

This low visual/Sérsic agreement is suggestive of potential inaccuracies
in the Sérsic classifications, especially at the higher redshifts
of our sample. While this could indeed be indicative of inaccuracies
in our visual classifications, we expect this is not the case, particularly
in light of the uniformly high agreement between our visual classifications
and the quantitative \citet{cassata2007} classifications. With this
high and uniform agreement, we conclude that using quantitative classifications
for the field sample while using visual classifications, instead of
Sérsic, for the cluster sample is appropriate. We have, however, run
the analysis both ways and have verified that none of the major qualitative
results depend on this choice. Our final sample of visually classified
isolated cluster ETGs (LTGs) contains 51 (61) galaxies, and the breakdown
by cluster is listed in Table\ \ref{tab:clusterList}.

\section{Analysis}

\label{sec:Analysis}

\begin{figure*}[!t]
\begin{centering}
\includegraphics{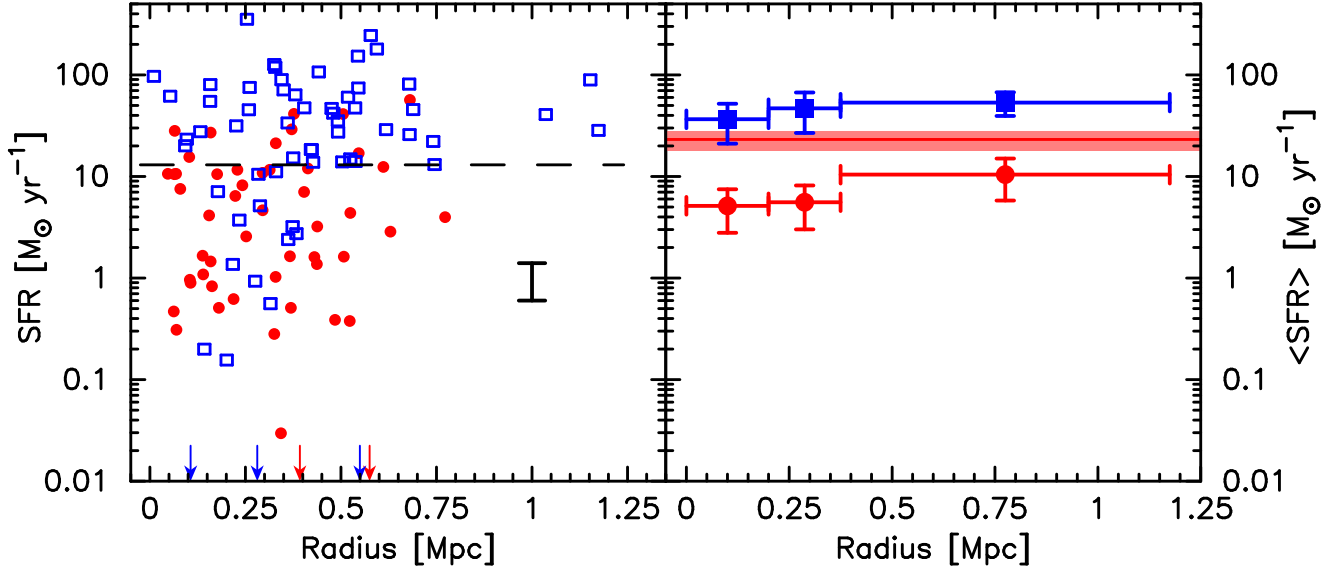}
\par\end{centering}

\protect\caption{\textit{Left panel}: SFR versus clustercentric radius for isolated
cluster ETGs (red filled circles), and LTGs (blue open squares). The
horizontal dashed line is the 1 $\sigma$ SFR detection level (13
\msun\ yr\textbf{$^{-1}$}), and the large error bar represents the
40\% systematic uncertainty in the SFRs. The arrows indicate the location
of cluster members undetected in MIPS (SFR $=0$). SFRs of isolated
cluster galaxies---of all morphological types---are weakly correlated
with projected radius. \textit{Right panel}: Mean SFR versus clustercentric
radius for isolated ETGs (red circles) and LTGs (blue squares), with
bin widths depicted by the horizontal error bars. In each bin, the
median SFR of non-detections is assigned to each undetected galaxy
and included in the mean. The vertical error bars show the quadrature
sum of bootstrap resampling and Poisson error. The solid horizontal
red line (shaded region) is the mean SFR (error) for isolated field
ETGs. On average, cluster ETGs are forming stars at 14\% the rate
of cluster LTGs, and nearly 30\% that of field ETGs. \label{fig:rad_SFR_meanSFR}}
\end{figure*}

\subsection{Star Formation Rate vs.\ Radius}

All the isolated galaxies in our sample are robustly detected in optical
and IRAC imaging, and our MIPS $24$ $\mu$m fluxes are measured for
\textit{all} sources using these positional priors. The resulting
SFRs are thus physically meaningful down to very low significances,
albeit with large uncertainties. 

In the left panel of Figure\ \ref{fig:rad_SFR_meanSFR} we plot SFR
versus projected clustercentric radius for isolated cluster ETGs (red
filled circles) and LTGs (blue open squares). Galaxies plotted below
the horizontal dashed line have SFRs below our 1 $\sigma$ depth of
13 \msun\ yr$^{-1}$. The large error bar shows the systematic error
in the SFR, which we take to be 40\%, based on a comparison between
24 $\mu$m and \herschel\ SFR measurements over $z=0$--1.5 \citep{elbaz2010}.

On their own, the ETG and LTG samples show little-to-no radial dependence
in the SFRs. However, when considering all isolated cluster galaxies,
we do find a weak correlation (Spearman's $r_{s}=0.26\pm0.06$ at
the 99\% confidence level) between SFR and clustercentric radius.

As can be seen in the left panel of Figure\ \ref{fig:rad_SFR_meanSFR},
we are largely limited to radii less than 0.75 Mpc due to the small
footprint of WFC3. We are able to probe beyond 1 Mpc in J1432.4+3250
due to two adjacent pointings on this cluster. We have verified our
results are unchanged if we limit our analysis to the well-sampled
region below 0.75 Mpc.

\subsection{Mean Star Formation Rate}

\label{sub:meanSFR}

To explore the effect of environment on the SFRs of both isolated
ETGs and LTGs, we show in the right panel of Figure\ \ref{fig:rad_SFR_meanSFR}
the mean SFR, $\left\langle SFR\right\rangle $, as a function of
projected clustercentric radius. ETGs are plotted as red circles and
LTGs as blue squares. We separate the galaxies into three non-overlapping
annuli, selecting the radial bins such that the S/N in each is approximately
equal. From inner to outer, the bin sizes are 200, 175, and 800 kpc,
respectively. The errors in each bin are calculated from the quadrature
sum of bootstrap resampling (1000 samples, with replacement) and simple
Poisson errors. We also show the mean SFR of our comparison sample
of isolated field ETGs with the red horizontal line. The 68\% error
in the mean is given as the shaded region.

In computing the mean SFRs, objects with individual SFRs below 13
\msun\ yr$^{-1}$ were assigned the median value of all such objects
in the bin. This is the catalog-space equivalent of median stacking;
from inner to outer annuli, the median SFRs of these ETGs (LTGs) are
1.5, 2.6, and 2.2 (0.2, 2.4, and 1.4) \msun\ yr$^{-1}$. We have
verified that none of our main results change even in the extreme
case of setting the SFRs of all such $<1$ $\sigma$ SFRs to zero.

We find mean SFRs of $5\pm2$, $6\pm3$, and $10\pm5$ \msun\ yr$^{-1}$
for our cluster ETG sample from the inner to outer annulus, respectively.
With mean SFRs of $40\pm20$, $50\pm30$, and $40\pm10$ \msun\ yr$^{-1}$
over the same range, the LTGs have mean SFRs $\sim$5 to 8 times higher
at all radii. Both ETGs and LTGs show some decrease in mean SFR at
small radii, although this is not statistically significant given
the large errors. Averaging over all radii we find a mean SFR of $7\pm2$
\msun\ yr$^{-1}$ for cluster ETGs, which is a factor of 7 lower
than for cluster LTGs ($\left\langle SFR\right\rangle =50\pm10$ \msun\
yr$^{-1}$). While the uncertainty in our SFRs is too large to determine
any radial dependence, we also find that cluster ETGs show suppressed
star formation activity relative to field ETGs, which have $\left\langle SFR\right\rangle =23\pm5$
\msun\ yr$^{-1}$.

These results show that although cluster ETGs have SFRs that are fairly
quenched relative to both their field analogs and the remainder of
the isolated cluster population, they still contribute 12\% of the
vigorous star formation observed amongst the isolated galaxies in
these clusters.

\subsection{Fraction of Star-forming Galaxies}

\label{sub:fractionSFR}

\begin{figure}[!t]
\begin{centering}
\includegraphics{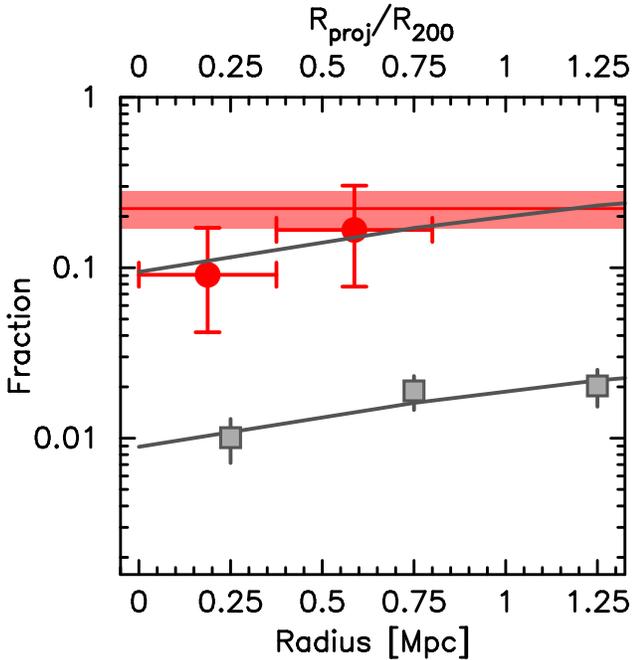}
\par\end{centering}

\protect\caption{Fraction of star-forming galaxies versus projected radius for high-redshift
isolated cluster ETGs (red circles) and low-redshift cluster galaxies
from C11 (gray squares and error bars). For cluster ETGs, the size
of the horizontal error bars represents the bin widths, and the vertical
error bars represent the binomial error in our fractions. The lower
axis corresponds to the data from this work; the upper from C11. The
lower gray curve is a least squares fit to all six points from Figure\ 4
of C11 (extending to $3R_{\textrm{proj}}/R_{200}$), while the upper
gray curve is the same fit, shifted up by a factor of 10.6. The red
horizontal line (shaded region) shows the fraction (error) for the
comparison high-redshift isolated field ETGs. High-redshift cluster
ETGs have star-forming fractions at least an order of magnitude higher
than local cluster galaxies, and two times lower than field ETGs in
the same redshift range.\label{fig:fractionSFR}}
\end{figure}

In Figure\ \ref{fig:fractionSFR}, we plot $f_{\textrm{SF}}$, the
fraction of star-forming isolated cluster ETGs (red circles) as a
function of clustercentric radius. We conservatively limit this measurement
to members with SFRs of at least 26 \msun\ yr$^{-1}$, above our
2 $\sigma$ detection level. We use only two annuli due to the relatively
small size of our sample, and use the binomial error in the fraction
as our total error. The radial bins and error ranges are shown by
the horizontal and vertical error bars, respectively. The horizontal
red line shows the fraction of star-forming isolated field ETGs, with
the binomial error in the fraction shown by the shaded region. The
gray points and error bars show the fraction of local ($z\lesssim0.1$)
star-forming ($L_{\textrm{IR}}>4.7\times10^{10}$ \lsun) cluster
galaxies versus projected radius, from \citet[hereafter C11]{chung2011},
who studied 69 low-redshift clusters with total dynamical masses in
the range $\sim\left(1-7\right)\times10^{14}$ \msun, determined
by using caustic infall patterns \citep{rines2006}, and selecting
only galaxies brighter than $M_{r}=-20.3$. The lower x-axis corresponds
to the projected clustercentric radius for this work, while the upper
x-axis corresponds to the projected $R_{200}$-normalized radius from
C11. Based on the X-ray, weak lensing and dynamical masses that have
been measured for a subset of the Bo\"otes clusters \citep{brodwin2011,jee2011},
as well as on a clustering analysis of the full ISCS sample \citep{brodwin2007},
our $z>1$ ISCS clusters have halo masses in the range $\sim$$(0.8-2)\times10^{14}$
\msun, and virial radii of $\sim$1 Mpc. Therefore the upper and
lower axes in Figure\ \ref{fig:fractionSFR} are approximately equivalent.
While the median mass of the C11 clusters is larger than that of the
Bo\"otes clusters, the latter will grow in mass in the $\sim$8--9
Gyr to the present epoch.

Averaging over all radii, we find that $12{}_{-5}^{+6}\%$ of isolated
cluster ETGs are star-forming. It is clear that the star-forming fraction
is substantially higher at $z>1.0$ than it is locally. We quantify
this difference by first fitting a least squares curve to all six
C11 points (while only the first three points are shown in Fig.\ \ref{fig:fractionSFR},
the C11 measurements extend to $3R_{\textrm{proj}}/R_{200}$), then
using $\chi^{2}$ minimization to determine that a simple scaling
factor of 10.6 provides an excellent fit to our data. In both radial
bins, we find that the fraction of star-forming ETGs in our sample
is approximately an order of magnitude higher than for local cluster
galaxies of all types.

Due to the very low redshift of the clusters in C11 ($z\lesssim0.1$),
their minimum cutoff for star-forming galaxies ($L_{\textrm{IR}}>4.7\times10^{10}$
\lsun) is $\sim3$ times lower than our 2 $\sigma$ level of $\sim$$1.5\times10^{11}$
\lsun. Using the published $L_{\textrm{IR}}$ values from C11, we
find that six of the 109 cluster galaxies (within $3R_{\textrm{proj}}/R_{200}$)
considered star-forming in C11 have $L_{\textrm{IR}}>1.5\times10^{11}$
\lsun, which is only $\sim$0.1\% of their cluster member sample,
a factor of $\sim$18 lower than with their SFR cut, and $\sim$100
times lower than our cluster ETG fraction.

Within the errors, we do not have enough evidence to determine whether
there is a radial trend in our ETGs, though the shifted radial profile
of C11, indicated by the upper curve, is clearly consistent with our
data. Also, the $f_{\textrm{SF}}$ of field ETGs ($0.22{}_{-0.05}^{+0.06}$)
is approximately twice that of the overall fraction for cluster ETGs,
implying an environmental dependence on isolated ETG star formation.

\subsection{Specific Star Formation Rate}

\label{sub:sSFR}

\begin{figure*}[!]
\begin{centering}
\includegraphics{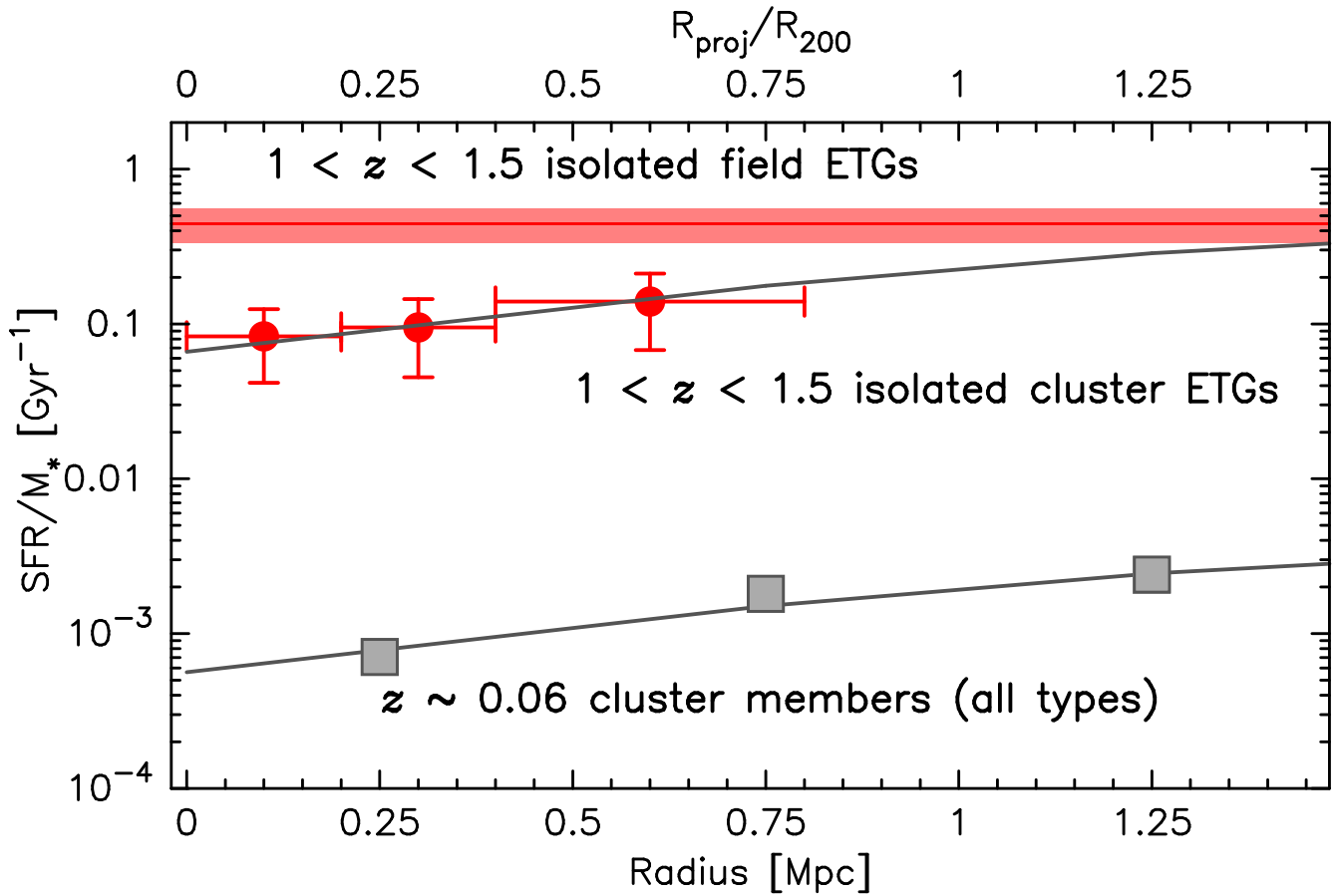}
\par\end{centering}

\protect\caption{Specific SFR versus radius for high-redshift isolated cluster ETGs
(red filled circles) and low-redshift cluster galaxies from C11 (gray
filled squares), with the same x-axes as Figure\ \ref{fig:fractionSFR}
(§\ref{sub:fractionSFR}). The vertical error bars represent the bootstrapping
and Poisson error in our sSFRs, and the horizontal error bars show
the size of each bin. The lower gray curve is a least squares fit
to the C11 points (all six points from their Fig.\ 3, extending to
$3R_{\textrm{proj}}/R_{200}$), while the upper gray curve is the
same fit shifted up by a factor of 120. The red horizontal line (shaded
region) shows the sSFR (error) for the comparison high-redshift isolated
field ETGs. The sSFR of high-redshift cluster ETGs is four times lower
than similar redshift field ETGs, yet more than two orders of magnitude
larger than low-redshift cluster galaxies of all types. While we place
no lower limit on the $L_{\textrm{IR}}$ of high-redshift cluster
ETGs, C11 only measured sSFR for galaxies with $L_{\textrm{IR}}>4.7\times10^{10}$
\lsun\ (their star-forming cut; see §\ref{sub:fractionSFR}). The
sSFR of C11's cluster galaxies would be lower if no cut was imposed,
making the factor of 120 we find here a lower limit. \label{fig:sSFR}}
\end{figure*}

We next explore the specific star formation rate (sSFR), defined as
the sum of the SFRs divided by the sum of the stellar masses, in each
radius bin. In normalizing the SFR of a galaxy to its mass, the sSFR
allows us to explore the relative efficiency with which it converts
its cold gas into stars.

In Figure\ \ref{fig:sSFR} we plot the sSFR versus radius for both
our cluster ETGs (red circles and error bars) and for the low-redshift,
star-forming, cluster galaxies of C11 (gray points). The radial binning
and error ranges for our points are calculated as in Figure\ \ref{fig:rad_SFR_meanSFR}.
We show the sSFR and similarly calculated error of field ETGs as the
red horizontal line and shaded region, respectively.

As in §\ref{sub:fractionSFR}, we fit a best-fit curve to the C11
points, then determine the shift in amplitude required to match our
cluster ETG points. We find that scaling the curve by a factor of
$10^{2.07}$, also shown in the figure, provides a good fit to our
high-redshift sSFR measurements. High-redshift cluster ETGs are forming
stars at a rate 120 times higher than local cluster galaxies.

C11 calculated sSFR for star-forming ($L_{\textrm{IR}}>4.7\times10^{10}$
\lsun) galaxies, while we place no such constraint on either high-redshift
ETG sample plotted in Figure\ \ref{fig:sSFR}. Additionally, the
morphological mix of the low redshift cluster galaxies to which we
are comparing is unclear. Although clusters in the local Universe
are primarily inhabited by early-type, ``red and dead'' galaxies
\citep{oemler1974,dressler1980}, the star-forming subset detected
by C11 may be preferentially drawn from the small fraction of late-type
members or from recently accreted field galaxies. Correcting for such
LTG contamination, and removing the $L_{\textrm{IR}}$ limit, would
lower the sSFR in the low redshift sample, and hence make the evolution
over this redshift range even more dramatic.

Averaging across all radii, high-redshift cluster ETGs have an sSFR
of $0.10\pm0.03$ Gyr$^{-1}$, which is a factor of 4 lower than the
sSFR of the field ETG sample ($0.4\pm0.1$ Gyr$^{-1}$). While we
cannot definitively determine whether there is a radial trend in cluster
ETG sSFR, this drop relative to ETGs in the field is further evidence
for the environmental dependence of the star formation of high-redshift
isolated ETGs, as shown in the previous two sections.

What is not immediately clear is how much of this offset between cluster
and field ETG sSFR is due to the difference in the distribution of
stellar masses and how much is due to star formation activity. If
the mass distribution of field and cluster ETGs was identical, we
could conclude that the large difference in sSFR between the two samples
was due entirely to star formation activity. However, even though
the minimum stellar mass allowed in each sample is the same ($1.3\times10^{10}$
\msun), the mean mass of cluster ETGs, $7.0\times10^{10}$ \msun,
is 1.4 times larger than that of field ETGs ($5.2\times10^{10}$ \msun).
Clearly, the lower sSFR of cluster ETGs must be due, at least in part,
to their higher stellar masses. To quantify this impact, we temporarily
replace the stellar mass of each field ETG with the mean stellar mass
of cluster ETGs. The sSFR of field early-types is now 0.3 Gyr$^{-1}$,
still a factor of 3 larger than for cluster ETGs. This value is 75\%
of the actual field ETG sSFR, which implies that only 25\% of the
offset is due to ETG stellar masses, and the bulk of the difference
is from the field-relative quenching of cluster ETGs.

\section{Discussion}

\label{Sec:Discussion}

\subsection{Comparisons with Related ISCS Work}

\label{sub:ISCS_Comparisons}

Two related studies (B13, and A14) identified high levels of star
formation activity in a superset of the ISCS clusters studied in this
work. However, these studies could not isolate the ETGs due to a lack
of morphological information. In the present work, which benefits
from the high-resolution imaging of \hst, we are able to expand upon
their results by analyzing the star formation properties of isolated,
massive ETGs, and comparing them to high-redshift, massive, isolated
ETGs in the field and to galaxies in low-redshift clusters.

\subsubsection{Comparison to B13}

\begin{figure}[!t]
\begin{centering}
\includegraphics{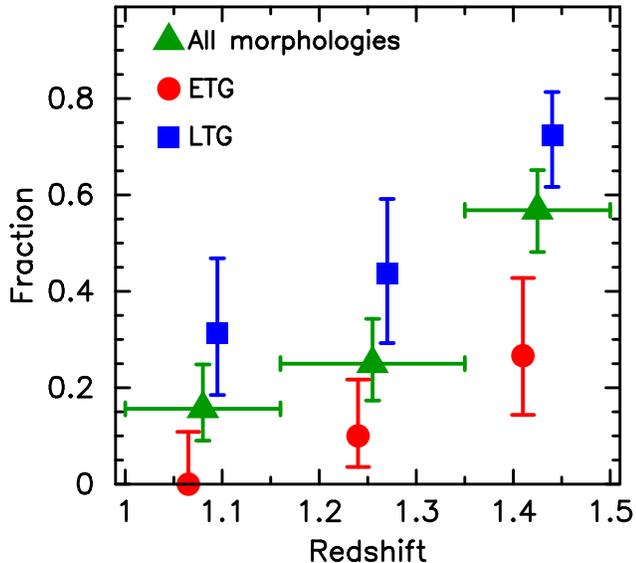}
\par\end{centering}

\protect\caption{Fraction of isolated, high-redshift star-forming (SFR $\geq26$ \msun\
yr$^{-1}$) cluster galaxies (see legend for symbol types) as a function
of redshift. Vertical error bars show the binomial error in the fractions,
while the horizontal error bars show the redshift binning for all
three sets of cluster galaxies. The significant drop in star-forming
fraction from 57\% to 25\% exhibited by all cluster galaxies (green
triangles) at $z\sim1.4$ is in good agreement with the redshift of
transition---away from vigorous star formation---observed by B13.
\label{fig:z_fractionSFR}}
\end{figure}

B13 measured the star formation activity in 16 ISCS galaxy clusters
at $1.0<z<1.5$, including the 11 clusters studied in this work. Their
large sample enabled them to finely bin their data as functions of
both redshift and radius. The morphology and isolation cuts in the
present work result in a relatively small sample size that precludes
a similar analysis, but instead permits an investigation of physically
interesting subsamples of these cluster members.

Both B13 and this work used the same 24 $\mu$m \spitzer\ photometry,
so the measurements of star formation activity should be consistent,
despite the sample size difference. To test this, we first compared,
as a function of redshift, the fraction of isolated star-forming galaxies
of \textit{all} types (by combining ETGs and LTGs) from this work
to the fraction of star-forming galaxies from B13 within 1 Mpc. We
used their redshift binning ($1<z<1.2$, $1.2<z<1.37$, and $1.37<z<1.5$)
and adopted their ($\textrm{S/N}\geq4$) flux limit, corresponding
to $\textrm{SFR}\gtrsim47$ \msun\ yr$^{-1}$. Our results are in
good agreement with those reported in B13.

Probing to lower star formation rates, we plot in Figure\ \ref{fig:z_fractionSFR}
the fraction of isolated star-forming cluster galaxies down to our
full 2 $\sigma$ $\textrm{SFR}\geq26$ \msun\ yr$^{-1}$ limit. We
bin the galaxies ($1<z<1.16$, $1.16<z<1.35$, and $1.35<z<1.5$;
shown by the horizontal error bars) such that the binomial error in
each redshift bin is approximately equal for galaxies of all morphological
type (green filled triangles). ETGs are plotted as red circles and
LTGs as blue squares, slightly offset to the left and right, respectively.

57\% of all isolated cluster galaxies of are star-forming at $1.35<z<1.5$,
followed by a decrease to 25\% in the middle redshift bin. This result
is in good agreement with the transition redshift of $z\sim1.4$ found
by B13 between the era of vigorous star formation in high-redshift
clusters and the quenched epoch at later times. The LTG population
experiences a similar trend, with a decrease from 72\% to 44\%, over
the same range, while the ETG sample shows a milder downward trend
in star-forming fraction at $z\sim1.4$ (from 27\% to 10\%), though
it's formally consistent with being constant within the errors.

\subsubsection{Comparison to A14}

\label{sub:comp_A13}

A14 explored 274 ISCS clusters from $z=0.3$ to 1.5, including $\sim$100
over the same redshift range as our observations. By stacking 250
$\mu$m \herschel\ data, they were able to probe to mean $L_{\textrm{IR}}$
values almost an order of magnitude lower than our 1 $\sigma$ detection
limit. Although our sample is a subset of the A14 sample, the measurement
techniques---24 $\mu$m detections versus stacking at 250 $\mu$m---are
relatively independent. Here we compare some of our results with those
reported in A14.

In order to compare the two SFR measurements for our morphologically
selected sample, we first attempted to directly measure stacked 250
$\mu$m fluxes for our ETG and LTG samples. However, with the relatively
small sample size, and source contamination due to the large beam
size (18\farcs1, \citealt{swinyard2010}), the S/N was too low to
permit this measurement.

\begin{figure}[!t]
\begin{centering}
\includegraphics{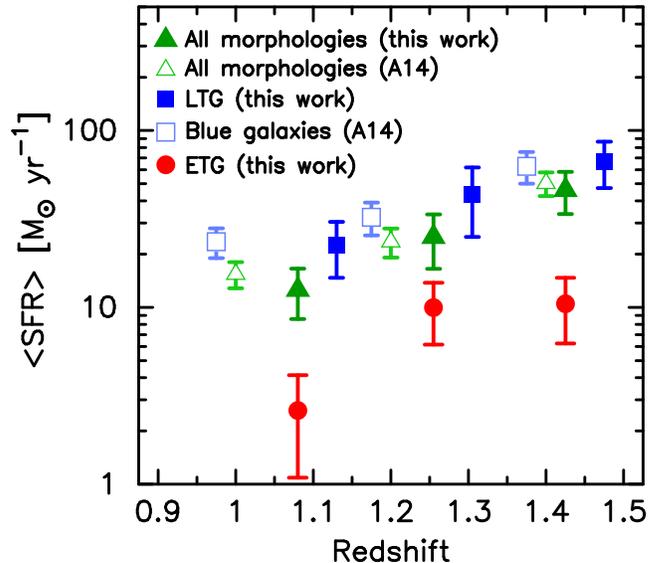}
\par\end{centering}

\protect\caption{Mean SFR versus redshift for our isolated cluster galaxies (filled
symbols, with the same binning as in Figure\ \ref{fig:z_fractionSFR})
and for cluster galaxies from A14 (open symbols), with all galaxies
having clustercentric radius $<$ 1 Mpc.\textit{ }Our errors are calculated
as the quadrature sum of bootstrapping and simple Poisson errors.
The two different measurement methods used for calculating these SFRs---24
$\mu$m detections versus stacking at 250 $\mu$m---show consistent
results. \label{fig:meanSFR_Alberts_Fit}}
\end{figure}

In Figure\ \ref{fig:meanSFR_Alberts_Fit} we compare mean SFR as
a function of redshift, derived from 24 $\mu$m and 250 $\mu$m measurements.
To be consistent with the radial selection in A14, we only plot galaxies
with clustercentric radii $<1$ Mpc. We plot our isolated cluster
galaxies as the filled points, using the same binning as in Figure\ \ref{fig:z_fractionSFR}.
The filled green triangles represent galaxies of all morphologies,
while the open triangles show the mean SFR of galaxies of all types
from A14. Despite the order of magnitude difference in observed wavelength,
and the very different measurement methodologies, the results are
in excellent agreement for galaxies of all morphological types.

While unable to visually or quantitatively determine morphologies,
A14 matched each galaxy in their sample against seven \citet{polletta2007}
templates representing different morphologies. Using this template
fitting as a proxy for color, A14 selected galaxies that were best
fit by late-type templates as ``blue'' (star-forming) galaxies.
We plot the mean SFR of these galaxies as the light blue open squares,
and compare them to our isolated cluster LTGs (blue filled squares).
We again find that there is excellent agreement between the 250 $\mu$m-derived
SFRs in A14 and the 24 $\mu$m-derived SFRs in this work.

Recent work suggests that SFRs based on MIR observations may be overestimated
for galaxies with evolved stellar populations. For example, thermally
pulsing asymptotic giant branch C stars may emit up to half their
light in the MIR \citep{kelson2010}. \citet{utomo2014} suggest,
for MIPS SFRs, up to a two order of magnitude overestimation of the
star formation of high $L_{\textrm{IR}}$ galaxies with evolved stellar
populations. 

The A14 results are based on 250 $\mu$m \herschel\ stacking, which
primarily probes the cold dust regime, and are largely immune to the
effects of evolved stellar populations that typically impact MIR wavelengths.
If our ETG SFRs were being overestimated to the extent suggested by
\citet{utomo2014}, one would expect that the star formation of our
\textit{entire} sample would be overestimated, since ETGs make up
approximately half of cluster galaxies. If this was the case, the
mean SFR of \textit{all} isolated cluster galaxies in Figure\ \ref{fig:meanSFR_Alberts_Fit}
would likely be higher than the mean SFR from A14. Due to the consistency
between our results and those of A14 for galaxies of all morphologies,
we surmise that our early-type population is not strongly impacted
by this effect.

\subsection{Star Formation in High-Redshift Cluster ETGs}

\label{sub:SF_high-z_ETGs}

Low-redshift ETGs, particularly those in clusters, have quiescent,
old stellar populations. A key issue in the evolution of galaxy populations
in clusters is determining the nature of the star formation history
of these ``red-and-dead'' galaxies. Specifically, when did present-day
massive cluster ETGs experience their last major burst of star formation?
Determining the epoch during which these galaxies experienced such
a burst can provide constraints on when cluster galaxies experienced
their last phase of gas-rich major merging.

We find that the fraction of star-forming galaxies is an order of
magnitude larger for our $1<z<1.5$ cluster ETGs than for local cluster
galaxies of \textit{all} morphologies (C11). Measurements of star-forming
fractions in nearby clusters, such as those by C11, necessarily include
large contributions from late-type galaxies as ETGs typically have
SFRs below the survey limits. The increase we find is therefore a
lower limit to the evolution between nearby and $z>1$ cluster ETGs.

Even more striking is the comparison of the mass-normalized star formation
rates between these two galaxy populations. With sSFRs more than two
orders of magnitude higher than cluster galaxies in the local universe,
our high-redshift cluster ETGs have significantly more ongoing star
formation activity per unit stellar mass. This dramatic evolution
would be even more extreme if C11 had measured the sSFRs of all galaxies
above a fixed mass limit, rather than just those above their star
formation detection limit.

An interesting result stemming from the comparisons in §\ref{sub:ISCS_Comparisons}
is that despite the decline in the fraction of star-forming ($\textrm{SFR}\geq26$
\msun\ yr$^{-1}$) cluster ETGs at $z\sim1.4$ seen in Figure\ $\ref{fig:z_fractionSFR}$,
this population has mean SFRs that are roughly constant across this
period. Specifically, although we see that the star-forming fraction
drops $\sim$17 percentage points (albeit with very large scatter)
at this redshift, their mean SFRs (shown in Figure\ $\ref{fig:meanSFR_Alberts_Fit}$)
remain relatively constant from $z\sim1.4\rightarrow1.25$. One potential
conclusion from this is that while a significant quantity of early-type
galaxies are being quenched, there must be some mechanism that is
\textit{enhancing} the star formation activity of the remaining star-forming
ETGs.

\begin{figure}[!t]
\begin{centering}
\includegraphics{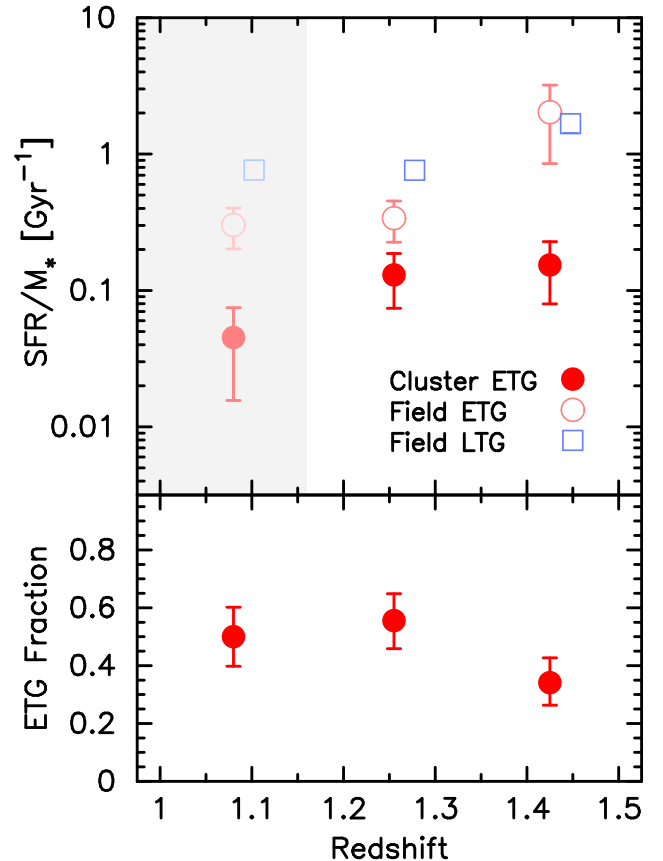}
\par\end{centering}

\protect\caption{\textit{Upper panel}: Specific SFR as a function of redshift for ETGs
in clusters (red filled circles) and in the field (pink open circles),
and for field LTGs (light blue open squares). Errors are calculated
as the quadrature sum of bootstrap resampling and simple Poisson errors,
as in §\ref{sub:sSFR}. We shade the lowest redshift bin to guide
the eye to two higher bins. Field galaxies show potentially strong
sSFR evolution over these two bins around $z\sim1.4$, while cluster
ETGs show little-to-no evolution over the same range. \textit{Lower
panel}: Fraction of cluster galaxies that are ETGs as a function of
redshift. \label{fig:z_sSFR_isolated}}
\end{figure}

If the above results are not a product of environment, we would expect
to see ETGs in both the cluster \textit{and} the field have a similar
lack of SFR evolution. To determine whether this is the case, we plot
in the upper panel of Figure\ \ref{fig:z_sSFR_isolated} the sSFR
for cluster and field galaxies, using the redshift binning from Figures\ \ref{fig:z_fractionSFR}
and \ref{fig:meanSFR_Alberts_Fit}. As we are primarily concerned
with potential evolution around $z\sim1.4$, we focus on the two higher
redshift bins, shading over the lowest bin. We plot cluster and field
ETGs with the filled red and open pink circles, respectively. Field
LTGs are represented by the open blue squares. Errors are from the
quadrature sum of bootstrap resampling and simple Poisson errors.

The sSFR of cluster ETGs is approximately the same in the two highest
redshift bins. Given the large uncertainties, though, we can make
no conclusions on whether a trend exists. However, we do find that
relative to field galaxies, the sSFR of cluster ETGs shows a weak---or
perhaps even no---dependence on redshift from the highest to the middle
redshift bin. This evidence suggests that any potential enhancement
of cluster ETG star formation may be environmentally dependent.

In the lower panel of Figure\ \ref{fig:z_sSFR_isolated} we plot
the fraction of isolated cluster galaxies that we classified as ETGs,
using the same redshift bins as above. We find that the fraction of
ETGs increases from 34\% to 56\% from $z\sim1.4\rightarrow1.25$,
which suggests that new ETGs are being formed during this period.

It should be noted that there are processes other than major merging
that may potentially play a role in forming new ETGs, or shaping existing
early-types \citep[e.g.,][]{kaviraj2013}. Violent disk instability
can cause late-type systems to lose their disks through turbulence,
forming compact gas-rich ``blue nuggets'' with gas inflows similar
to those generated by wet mergers \citep{dekel2014}. However, the
cold streams \citep{keres2005,dekel2006} that feed disks in this
model are only important in regions with low galaxy density \citep{keres2005},
and likely not a significant factor in the hot halo environment of
ISCS clusters.

\citet{strazzullo2010} found a large fraction of quenched, compact
ETGs in the $z=1.39$ cluster XMMU J2235, suggesting that minor---and
likely dry---mergers can increase the size of such galaxies over later
epochs, without drastically altering their star formation activity.
However, XMMU J2235 is a very massive cluster \citep[$\sim$$7\times10^{14}$ \msun,][]{jee2009,rosati2009},
where major merger activity has likely ceased, and a factor of at
least a few times more massive than the clusters studied in this work.
As such, the mechanism suggested by \citet{strazzullo2010} is not
likely \textit{currently} playing a role in ISCS clusters---especially
at $z\gtrsim1.16$---when considering the star formation activity
shown above.

From $z\sim1.4\rightarrow1.25$, the fraction of cluster ETGs that
are star-forming drops from 27\% to 10\%. It is likely that a substantial
number of cluster ETGs formed before $z=1.5$, and that they make
up a significant portion of this large subset of quenched ETGs that
we find in ISCS clusters. However, from $1.35<z<1.5$ to $1.16<z<1.35$,
there is a $\sim$20 percentage point increase in the fraction of
cluster galaxies that are morphologically early-type; some portion
of the remaining star-forming ETGs in this epoch are likely recent
byproducts of major mergers. Furthermore, cluster ETG mean and specific
SFRs are roughly constant over this period. These ETGs have not yet
had sufficient time for their star formation to be quenched, implying
that their progenitors' mergers occurred relatively recently.

Moving into the lower redshift bin ($1<z<1.16$), the star-forming
ETG fraction falls to 0\%, and their sSFR and mean SFR are both quenched
(by factors of 3 and 4, respectively), while the ETG fraction remains
relatively constant. Given our assumptions above about gas-rich major
mergers, the dearth of star-forming galaxies, and overall lack of
star formation activity seen in this epoch, at $1<z<1.16$, suggests
we are seeing the quenching of ETGs due to post-merger AGN activity.

\section{Conclusions}

\label{Sec:Conclusions}

We have used a sample of 11 high-redshift ($1.0<z<1.5$), IR-selected
ISCS galaxy clusters to investigate the star formation properties
of isolated, early-type galaxies. After conservatively removing AGNs
through X-ray and IR criteria, we visually inspected our sample using
high-resolution \hst\ imaging, separating our galaxy sample into
two coarse morphological bins: ETGs and LTGs. We used deep 24 $\mu$m
imaging from \spitzer\ to measure the obscured SFRs, excluding galaxies
for which we could not rule out contamination from nearby neighbors.

We compared the star formation of the cluster ETG sample with low-redshift
cluster galaxies, finding an order of magnitude larger fraction of
star-forming galaxies, and a greater than two order of magnitude larger
sSFR for our high-redshift cluster ETGs. Averaging across our entire
cluster ETG sample, we find that 12\% are still experiencing relatively
enhanced ($\textrm{SFR}>26$ \msun\ yr$^{-1}$) star formation activity.

By comparing the mean SFR of ETGs with LTGs in ISCS clusters, we found
that despite their enhanced star formation relative to low-redshift
cluster galaxies, high-redshift cluster early-types have substantially
less star formation activity relative to the rest of the isolated
cluster population. However, averaging across all cluster radii, ETGs
still contribute 12\% of the significant star formation activity observed
in these clusters.

Due to our relatively small sample size, we were unable to detect
the radial dependence in star formation activity reported by B13.
However, we found that our cluster ETGs are quenched relative to a
comparison sample of field ETGs, by a factor of 3 in mean SFR, and
4 in sSFR. Even when ``correcting'' the sSFR of field ETGs for their
factor 1.4 lower mean stellar masses, we find that it only accounts
for 25\% of the difference between the field ETG sSFR and the significantly
quenched cluster ETG sSFR.

We then used the conservative IR luminosity cut from B13 to compare
the fraction of star-forming galaxies, $f_{\textrm{SF}}$, with their
results, finding that our measurements in isolated galaxies agreed
with the B13 measurements of all cluster galaxies in these $z>1$
clusters. We also found that our mean SFR measurements correlated
well with those of A14, who measured SFR using stacked 250 $\mu$m
\herschel\ flux.

We used our 2 $\sigma$ SFR detection limit (26 \msun\ yr$^{-1}$)
to explore the $f_{\textrm{SF}}$ evolution from $z\sim1.5\rightarrow1.0$.
We considered cluster galaxies of all morphologies and found that
while 57\% are star-forming at $z\gtrsim1.4$, the fraction drops
to 25\% by $z\sim1.25$. This drop of more than 30 percentage points
suggests that the epoch of enhanced star formation in these clusters
is ending around $z\sim1.4$, a finding consistent with that first
reported by B13.

While the fraction of star-forming ETGs drops from 27\% at $z\sim1.4$
to 10\% by $z\sim1.25$, their mean and specific SFRs are largely
unchanged over this period. With a corresponding increase of $\sim$20
percentage points in the fraction of galaxies classified as early-type,
these results are consistent with a scenario where major gas-rich
mergers form new early-type galaxies, temporarily enhancing their
star formation activity.

A number of recent studies of the ISCS cluster population have presented
lines of evidence supporting the role of mergers in building the stellar
mass in these clusters. Specifically, the NIR luminosity function
evolution disagrees with a passive evolution model at $z\gtrsim1.3$
\citep{mancone2010,mancone2012}, galaxies are experiencing substantial
star formation (A14, B13, \citealt{zeimann2013}), young galaxies
are continuously migrating on the cluster red sequence (S12), and
an increase in AGN activity has been observed \citep{galametz2009,martini2013}.
This work helps to solidify the implications of these studies by demonstrating
the redshift dependence of the star formation rate and ETG fraction
that are consistent with the picture where massive ETGs are formed
in gas-rich major mergers.

\acknowledgements This work is based in part on observations made with the {\it{Spitzer Space Telescope}}, which is operated by the Jet Propulsion Laboratory, California Institute of Technology under a contract with NASA. Support for this work was provided by NASA through an award issued by JPL/Caltech. Support for \hst\ programs 10496, 11002, 11597, and 11663 were provided by NASA through a grant from the Space Telescope Science Institute, which is operated by the Association of Universities for Research in Astronomy, Inc., under NASA contract NAS 5-26555. This work is based in part on observations obtained with the {\it{Chandra X-ray Observatory}}, under contract SV4-74018, A31 with the Smithsonian Astrophysical Observatory which operates the {\it{Chandra X-ray Observatory}} for NASA.

We thank S. Chung for providing her data in digital form. We also appreciate the UltraVISTA and COSMOS collaborations making their data publicly available.

\bibliographystyle{apj}
\bibliography{bibtex}

\end{document}